\documentstyle[12pt]{article}
\begin{document}

\centerline {\bf Stochastic Resonance}

\vskip 0.5truecm
\centerline {M.I. Dykman$^1$, D.G. Luchinsky$^2$, R. Mannella$^3$,}
\centerline {\underbar {P.V.E. McClintock}$^4$, S.M. Soskin$^5$,
N.D. Stein$^4$ and N.G. Stocks$^6$}

\vskip 0.3truecm
\centerline {$^1$Department of Physics, Stanford University, Stanford, CA
94305, USA.}

\vskip 0.3truecm
\centerline {$^2$VNIIMS, Andreevskaya nab 2, 117965 Moscow, Russia.}

\vskip 0.3truecm
\centerline {$^3$Dipartimento di Fisica, Universit$\grave{\rm a}$ di Pisa,
Piazza Torricelli 2, 56100 Pisa, Italy.}

\vskip 0.3truecm
\centerline {$^4$School of Physics and Materials, Lancaster University,
Lancaster, LA1 4YB, UK.}

\vskip 0.3truecm
\centerline {$^5$Institute of Semiconductor Physics, pr. Nauki 45,
252038 Kiev, Ukraine.}

\vskip 0.3truecm
\centerline {$^6$Department of Engineering, Warwick University, Coventry,
CV4 7AL, UK.}

\begin{abstract}
Stochastic resonance (SR) - a counter-intuitive phenomenon in which
the signal due to a weak periodic force in a nonlinear system can be
{\it enhanced} by the addition of external noise - is reviewed.  A
theoretical approach based on linear response theory (LRT) is
described.  It is pointed out that, although the LRT theory of SR is by
definition restricted to the small signal limit, it possesses substantial
advantages in terms of simplicity, generality and predictive power.
The application of LRT to overdamped motion in a
bistable potential, the most commonly studied form of SR, is outlined.
Two new forms of SR, predicted on the basis of LRT and subsequently
observed in analogue electronic experiments, are described.
\end{abstract}

\vskip 0.5truecm
\noindent
{\bf 1. Introduction}

One of the most active current growth areas of nonlinear dynamics lies in
the relatively unexplored region separating the two major
divisions of the subject: that is, within the interface separating
\lq\lq deterministic" nonlinear dynamics, (e.g. Thompson and Stewart,
1986), where externally applied forces are
precisely known (e.g. periodic), from stochastic nonlinear dynamics
where the system under study fluctuates under the influence of a random
force (e.g. Moss and McClintock, 1989).  {\it Stochastic resonance} (SR), in
which the signal due to a weak periodic force in a nonlinear system can,
remarkably, be amplified by the addition of external noise, provides an example
of a phenomenon in this interface region.  It arises through a
tripartite interaction between nonlinearity, fluctuations and a
periodic force, and it cannot occur unless all three of
these features are simultaneously present.

The notion of SR was originally introduced (Nicolis, 1982; Benzi et al, 1982)
in relation to the earth's ice-age cycle.  The phenomenon
was subsequently demonstrated in an electronic circuit (Fauve and
Heslot, 1983) and in a ring laser (McNamara et al, 1988).
Following this latter paper, there has been a veritable explosion of
activity leading to the observation of SR or associated phenomena in
a wide variety of contexts, including passive optical systems (Dykman
et al, 1991), electron spin resonance (Gammaitoni et al, 1991a), sensory
neurons (Longtin et al, 1991) and a magneto-elastic strip (Spano et al,
1992).  These references are merely illustrative: a fuller bibliography
can be found within the proceedings of a recent conference on SR
(Moss, Bulsara and Shlesinger, 1993).

In this paper we introduce SR and set it within the context of classical
linear response theory (LRT).  We emphasize that the LRT perception
of the phenomenon is very general.  Not only does it provide a good
description of SR in systems with static bistable potentials (conventional
SR) but it also leads on naturally to the prediction of new forms of SR in
quite different kinds of systems: see Wiesenfeld (1993).  In Section 2 we
describe this LRT approach and in Section 3 we show how it may be applied to
conventional SR.  Sections 4 and 5 describe two quite new forms of SR -
associated with fluctuational transitions between coexisting periodic
attractors, and for underdamped nonlinear oscillators in the absence of
bistability - that were predicted on the basis of LRT and subsequently
observed in electronic experiments.  In Section 6 we summarise the results,
discuss future directions, and draw conclusions.

\vskip 0.5truecm
\noindent
{\bf 2. Linear response theory of stochastic resonance}

We shall define SR as an increase of the
amplitude of a periodic signal in a nonlinear system resulting from
the addition of external noise at the input; often, the signal/noise
ratio at the output will also increase, an effect that meets the
stricter definition of SR used by some authors.  In both cases, the signal
decreases again for sufficiently strong noise, giving rise to a resonance-like
curve when the amplitude is plotted against noise intensity, thereby accounting
for the terminology.

The theory of SR has been perceived as difficult, because of the need
to treat stochastic and periodic forces together in a highly nonlinear
system.  It has mostly been developed with the simplifying assumption
of a discrete two-state model (in the case of bistable systems) or, in the
case of continuous systems, has been based on an approximate or numerical
solution of the Fokker-Planck equation for a periodically driven system,
sometimes with contradictory results (Benzi et al 1982; Nicolis,
1982; Presilla et al, 1989; Gammaitoni et al, 1989; McNamara and
Wiesenfeld, 1989; Fox, 1989; Hu Gang, Nicolis and Nicolis, 1990; Jung
and Hanggi, 1990, 1991).

The alternative approach to SR introduced by Dykman et al
(1990a, 1990b), based on LRT, is quite different.  According to
LRT (see e.g. Landau and Lifshitz, 1980), if a system with
coordinate $q$ is driven by a weak force $A \cos \Omega t$,
a small periodic term $\delta \langle q (t) \rangle$ will appear in
the ensemble-averaged value of the coordinate, oscillating at the same
frequency $\Omega$

\begin{equation}
\delta \langle q (t) \rangle = a \cos (\Omega t + \phi), \quad A \to 0
\end{equation}

\begin{equation}
a = A \vert \chi (\Omega) \vert, \quad \phi = - {\rm arctan}
[{\rm Im} \chi (\Omega)/{\rm Re} \chi (\Omega)]
\end{equation}

\noindent
where $\chi (\Omega)$ is the {\it susceptibility} of the system.  The function
$\chi (\Omega)$ contains virtually everything needing to be known about the
response of the system to a weak driving force.  It gives both the
{\it amplitude} $a$ of the signal and its {\it phase lag} $\phi$ relative to
the driving force.  The occurrence of a delta-shaped spike at frequency
$\Omega$ in the spectral density of fluctuations (SDF), $Q (\omega)$, of the
system

\begin{equation}
Q (\omega) = \lim_{\tau \to \infty} (4 \pi \tau)^{-1} \vert \int^{\tau}_{-\tau}
dt q (t) \exp (i \omega t) \vert^2
\end{equation}

\noindent
follows immediately from (1) on account of the principle of the decay
of correlations

$$\langle q (t) q (t^{\prime}) \rangle \to \langle q (t) \rangle
\langle q (t^{\prime})\rangle \quad {\rm for} \quad \vert t - t^{\prime}
\vert \to \infty$$

\noindent
The {\it intensity} $\frac {1}{4} a^2$ (i.e. area) of the spike can be found
from (2).  Following McNamara et al (1988), the signal/noise ratio in SR is
often defined as the ratio $R$ of the area of the spike to the value
$Q^{(0)} (\Omega)$ of the SDF at frequency $\Omega$ but in the
absence of the driving force.  From (1) - (3), this quantity may be
expressed in terms of the susceptibility as

\begin{equation}
R = \frac{1}{4} A^2 \vert \chi (\Omega) \vert^2 / Q^{(0)} (\Omega)\quad (A \to
0)
\end{equation}

\noindent
Consequently, the evolution of $\chi (\Omega)$, or of $\chi (\Omega)$
and $Q^{(0)} (\Omega)$, with increasing noise intensity shows
immediately whether or not SR in the signal or in the signal/noise ratio,
respectively, is to be expected at frequency $\Omega$ in any given
system.

In the particular case of systems that are in thermal equilibrium, or
quasi-equilibrium, the susceptibility at frequency $\Omega$ can be
obtained very simply from the fluctuation dissipation relations
(Landau and Lifshitz, 1980),

$${\rm Re} \chi (\Omega) = \frac {2}{T} {\rm P} \int^{\infty}_{0} d \omega_1
Q^{(0)} (\omega_1) \omega_1^2 (\omega^2_1 - \Omega^2)^{-1}$$

\begin{equation}
{\rm Im} \chi (\Omega) = \frac {\pi \Omega}{T} Q^{(0)} (\Omega)
\end{equation}

\noindent
where P implies the Cauchy principal part, and $T$ is the temperature (noise
intensity) in energy units.  It is interesting to note that a knowledge
of $Q^{(0)} (\omega)$ and its evolution with $T$ is then sufficient in itself,
to predict whether or not the system in question will exhibit SR: this
would be true even where the underlying dynamics was unknown, and the
information about $Q^{(0)} (\omega)$ had been acquired by experiment.

\vskip 0.5truecm
\noindent
{\bf 3. Stochastic resonance in static double-well potentials}

The initial tests of the above ideas were performed (Dykman et al, 1990a,
1990b) through the measurement of SDFs and the investigation of SR in an
electronic model of the damped double-well Duffing oscillator,

\begin{equation}
\ddot q + 2 \Gamma \dot q + U^{\prime} (q) = A \cos \Omega t + f(t)
\end{equation}

$$U (q) = - \frac{1}{2} q^2 + \frac {1}{4} q^4, \quad \langle f (t) \rangle =
0, \quad \langle f (t) f(t^{\prime})\rangle = 4 \Gamma T \delta (t -
t^{\prime})$$

\noindent
The results are shown in Fig 1, where the scaled signal/noise ratio
$\tilde R$ = 6.51 $\times 10^{-4} R$ is plotted as a function of scaled
noise intensity $T/\Delta U$, $\Delta U$ ($= \frac {1}{4}$ for the potential
in (6)) being the height of the potential barrier between the wells.  The
square data points represent direct measurements of $\tilde R$, obtained from
the heights of the delta spikes in $Q (\omega)$; the crosses are also obtained
experimentally, but in a completely different way, from Equations (2), (4) and
(5) using measurements of $Q^{(0)} (\omega)$ in the absence of the periodic
force.  The fact that the agreement is excellent, within the experimental
error, without the use of any adjustable parameters, can be regarded as a
direct confirmation of the validity of the LRT perception of small signal SR.

For the particular parameters used for the measurements of Fig 1,
the magnitude of the rise in $\tilde R$ is relatively  modest;
much larger increases can be obtained for lower frequencies $\Omega$
and larger damping constants $\Gamma$.  Nonetheless, it is clear that
there is a range of $T/\Delta U$ within which $\tilde R$ rises with
increasing $T$, i.e. there is a manifestation of SR.

Of course, for a theory of SR, one would also need to be able to calculate
$Q^{(0)} (\omega)$, rather than having to measure it experimentally.
Although this has been done (Dykman et al, 1988) for the system (6), the
most detailed experimental and theoretical studies relate to the equivalent
overdamped system, which has been widely used as the standard system for
investigations of SR,

\begin{equation}
\dot q + U^{\prime} (q) = A \cos \Omega t + f (t)
\end{equation}

$$U (q) = - \frac {1}{2} q^2 + \frac {1}{4} q^4, \quad \langle f (t)
\rangle = 0, \quad \langle f (t) f(t^{\prime}) \rangle = 2D\delta (t
- t^{\prime})$$

\noindent
where $f (t)$ is now a zero-mean Gaussian noise of intensity $D$
(the way in which the overdamped limit is taken to obtain (7)
from (6) is discussed e.g. by Risken (1989)).  Like (6), (7) for $A$ = 0
is also a thermal equilibrium system so that, in order to find the
susceptibility $\chi (\omega)$, it is only necessary to calculate the SDF
$Q^{(0)} (\omega)$ in the absence of the periodic force for substitution
in the fluctuation dissipation relations (5) with $T$ replaced by $D$.  In the
limit of weak noise, $D \ll \Delta U$, both $Q^0 (\omega)$ and $\chi (\omega)$
can be obtained analytically (Dykman and Krivoglaz, 1979, 1984; Dykman et al
1989) as a sum of partial contributions from fluctuations about the equilibrium
positions $q_n$ and from interwell transitions,

\begin{equation}
Q^{(0)} (\omega) = \Sigma_{n=1,2}^{} w_n Q^{(0)}_n (\omega) + Q^{(0)}_{tr}
(\omega),
\,\,\,\,\chi (\omega) = \Sigma_{n=1,2}^{} w_n \chi_n (\omega) + \chi_{tr}
(\omega)
\end{equation}

\noindent
Here $w_n$ is the population of the nth stable state and, for the model (7),
$w_1 = w_2 = \frac {1}{2}$, $Q^{(0)}_1 (\omega) = Q_2^{(0)} (\omega)$
and $\chi_1 (\omega) = \chi_2 (\omega)$.  The SDF for the intrawell
vibrations $Q^{(0)}_n (\omega)$ is obtained by expanding $U (q)$
about the equilibrium position; $Q^{(0)}_{tr} (\omega)$ can be written down in
terms of the transition probabilities $W_{nm}^{(0)}$, defining the
probability of an $n \to m$ transition in the absence of periodic
forcing.

These calculations result in explicit analytic predictions for $R(D)$ and
$\phi (D)$.  The latter is of particular interest in view of the inconsistent
results of earlier calculations and experiments, with  (Nicolis, 1982; McNamara
and Wiesenfeld, 1989; Hu Gang et al 1990) on the one hand, and (Gammaitoni
et al, 1990, 1991 a, b) on the other.  An analogue electronic experiment was
performed (Dykman et al, 1992a) to measure $\phi (D)$ for comparison with
the LRT theoretical predictions, yielding the results shown by the data points
in the main section of Fig 2; the inset shows a plot of $R/A^2$ as a function
of $D$ in the range near the minimum where other theories fail.  In both
cases, the agreement between experiment and theory is very satisfactory,
providing further confirmation of the validity of the LRT approach to SR.
The dashed line shows the prediction of earlier (two-state)
theories (e.g. Nicolis, 1982; McNamara and Wiesenfeld, 1989) that do
not include the effect of intrawell motion.

\newpage
\noindent
{\bf 4. Stochastic resonance for periodic attractors}

It is clear from the above discussion that any system whose susceptibility
$\chi (\Omega)$ increases with noise intensity may be expected to display
SR when driven by a weak periodic force of frequency $\Omega$. Dykman and
Krivoglaz (1979) had noticed such an effect in the imaginary part of the
susceptibility of a periodically driven nonlinear oscillator with coexisting
periodic attractors.  It was therefore obvious that SR was to be expected in
systems of this kind, and that it would be likely to have some unusual and
characteristic features distinguishing it from conventional SR.  The new
phenomenon has been sought and recently found and investigated (Dykman et al,
1993 b, c) .  The results have implications for a large class of passive
optically bistable systems and, in particular, for optically bistable
microcavities.

The system that we consider is the nearly-resonantly-driven, underdamped,
single-well Duffing oscillator with additive noise,

\begin{equation}
\ddot q + 2 \Gamma \dot q + \omega^2_0 q + \gamma q^3 = F \cos (\omega_F
t) + f(t)
\end{equation}

$$\Gamma, \vert \delta \omega \vert \ll \omega_F, \quad \gamma \delta
\omega > 0, \quad \delta \omega = \omega_F - \omega_0, \quad \langle f (t)
\rangle = 0, \quad \langle f (t) f (t^{\prime}) \rangle = 4 \Gamma T \delta
(t - t^{\prime})$$

\noindent
Note that the force $F \cos (\omega_F t)$ is not very weak; neither is it so
strong that the system becomes chaotic or displays subharmonics.  Within a
certain parameter range, (9) is characterised by two coexisting periodic
attractors of different amplitude and phase.  Weak noise $f(t)$ causes
occasional transitions between them.  For appropriate noise intensity, these
transitions can become coherent on average with an additional weak
periodic trial force $A \cos (\Omega t + \phi)$ added to (9), provided
that $\Omega$ is close to $\omega_F$, leading to stochastic amplification,
i.e. SR.  It can be shown that the system responds strongly to the trial
force, not only at $\Omega$ but also at $\vert 2 \omega_F - \Omega \vert$;
the relevant susceptibilities can be calculated by an extension of the
Dykman and Krivoglaz (1979) theory.

Measurements of the signal/noise ratio $R$ in an analogue electronic
experiment (data points) are compared with the theoretical predictions
in Fig 3.  Although the results are very similar to those found
in conventional SR (Moss et al, 1993), it must be emphasized that SR for
periodic attractors also posesses a number of features that are entirely
different.  The most important of these are, first, that it is a high
frequency phenomenon.  Stochastic amplification takes place not at a
low frequency comparable to the inter-attractor hopping rate, as in
conventional SR, but at the much higher frequency $\Omega$ comparable
to $\omega_0$.  Secondly the stochastic enhancement of the
signal at the mirror-reflected frequency $\vert \Omega - 2 \omega_F \vert$
has no analogue in conventional SR.  Other differences, and the
relationship to phenomena in nonlinear optics, are discussed by Dykman et
al (1993 c).

\vskip 0.5truecm
\noindent
{\bf 5. Stochastic resonance in monostable systems}

Until recently, it was the almost universal assumption (Moss et al,
1993, and references therein) that the stochastic amplification of
SR could occur only as the result of nearly periodic fluctuational
transitions between coexisting attractors, corresponding to the minima of a
static bistable potential.  The high-frequency SR of Section 4 extends the
picture to encompass periodic attractors, but it still requires bistability.
The LRT picture of SR, however, does not involve any such requirement:
any system, bistable or otherwise, in which the susceptibility $\chi (\Omega)$
increases with noise intensity would, in view of (1), (2), be expected to
display SR.

One monostable system in which SR is to be anticipated on these grounds is the
underdamped, single-well, Duffing oscillator subject to a constant field

\begin{equation}
\ddot q + 2 \Gamma \dot q + U^{\prime} (q) = A \cos \Omega t + f(t)
\end{equation}

$$U (q) = \frac {1}{2} q^2 + \frac {1}{4} q^4 + Bq, \quad \Gamma \ll 1, \quad
\langle f (t) \rangle = 0, \quad \langle f (t) f(t^{\prime}) \rangle = 4
\Gamma T \delta (t-t^{\prime})$$

\noindent
which is known (Stocks et al, 1993a) to have extremely sharp
zero-dispersion peaks (ZDPs) in its SDFs provided
that $\vert B \vert > 8/(7)^{3/2}$ so that the variation of the
oscillator's eigenfrequency $\omega (E)$ with energy $E$ posesses
an extremum (Dykman et al, 1990c).  The ZDPs rise exponentially
fast with increasing noise intensity $T$.  Thus, because (13) for $A$ = 0 is a
system of the thermal equilibrium type, to which (5) is applicable,
$\chi (\Omega)$ may also be expected to rise extremely fast provided
that $\Omega$ is chosen to be in the close vicinity of the ZDP.

Experimental results (Stocks et al, 1993b) obtained from an analogue
electronic model of (10) with $B$ = 2 are shown by the circle data
of Fig 4(a).  The quantity plotted is the square of the stochastic
amplification factor

$$S (T) = a(T)/a(0)$$

\noindent
where $a(T)$ is the amplitude of the signal for noise intensity $T$.
The fact that $S(T)$ rises very rapidly (from the value of unity that it
would take in the absence of stochastic amplification) provides a clear
signature of SR.  The fuller curve is the LRT prediction, based on (1) and (5)
using the expressions for $Q^{(0)} (\omega)$ given by Dykman et al
(1990c).  It is in very satisfactory agreement with the data.  The phase
shift $\phi (T)$ has also been measured and calculated, as shown by the
circle data and associated curve in Fig 4(b).  Here, too, experiment and
LRT theory agree well.   A physically motivated discussion (together with
an explanation of the experimental and theoretical results obtained for
$B$ = 0 shown by the square data and associated curves) of this new form of
SR has been given by Stocks et al (1993b).  It can be demonstrated
(Stocks et al, 1992) on the basis of LRT that, for sufficiently small
$\Gamma$ in (10), substantial increases, not only in the signal, but
also in the signal/noise ratio $R$ are to be expected.

\vskip 0.5truecm
\noindent
{\bf 6. Conclusion}

We conclude that LRT provides a good description, not only of
conventional SR, but also of the other new forms of SR that can be
predicted on that basis.  These include SR for periodically modulated
noise (Dykman et al 1992b) which has not been considered here, as well
as the SR for periodic attractors and monostable systems discussed above.
Although LRT is, by definition, applicable only in the small signal limit,
in combination with the corresponding physical picture of SR it provides a
valuable clue as to the general type of behaviour to be expected
even for larger amplitudes of the trial force.

\newpage
\noindent
{\bf References}

Benzi, R., Parisi, G., Sutera, A. and Vulpiani, A. 1982, Stochastic
resonance in climatic change. {\it Tellus} {\bf 34}, 10-16.

Dykman, M.I., and Krivoglaz, M.A. 1979, Theory of fluctuational transitions
between stables states of a nonlinear oscillator. {\it Sov. Phys. JETP}
{\bf 50}, 30-37.

Dykman, M.I., and Krivoglaz, M.A. 1984, Theory of nonlinear oscillator
interacting with a medium, in {\it Soviet Physics Reviews}, ed Khalatnikov,
I.M., Harwood, New York, vol 5, pp 265-441.

Dykman, M.I., Mannella, R., McClintock, P.V.E., Moss, F. and Soskin, S.M.
1988, Spectral density of fluctuations of a double-well Duffing oscillator
driven by white noise. {\it Phys. Rev. A} {\bf 37}, 1303-1313.

Dykman, M.I., Krivoglaz, M.A. and Soskin, S.M. 1989, Transition
probabilities and spectral density of fluctuations of noise driven
bistable systems, in {\it Noise in Nonlinear Dynamical Systems}, ed.
Moss, F. and McClintock, P.V.E., Cambridge University Press, vol 2, pp 347-380.

Dykman, M.I., Mannella, R., McClintock, P.V.E., and Stocks, N.G.
1990a, Comment on stochastic resonance in bistable systems.
{\it Phys. Rev. Lett.} {\bf 65}, 2606.

Dykman, M.I., McClintock, P.V.E., Mannella, R. and Stocks, N.G.
1990b, Stochastic resonance in the linear and nonlinear responses of a
bistable system to a periodic field. {\it Sov. Phys. JETP Lett.} {\bf 52},
141-144.

Dykman, M.I., Mannella, R., McClintock, P.V.E., Soskin, S.M. and Stocks, N.G.
1990c, Noise-induced narrowing of peaks in the power spectra of
underdamped nonlinear oscillators. {\it Phys. Rev. A} {\bf 42},
7041-7049.

Dykman, M.I., Velikovich, A.L., Golubev, G.P., Luchinsky, D.G. and
Tsuprikov, S.V. 1991, Stochastic resonance in an all-optical passive bistable
system. {\it Sov. Phys. JETP Lett.} {\bf 53}, 193-197.

Dykman, M.I., Mannella, R., McClintock, P.V.E., and Stocks, N.G.
1992a, Phase shifts in stochastic resonance. {\it Phys. Rev. Lett.} {\bf 68},
2985-2988.

Dykman, M.I., Luchinsky, D.G., McClintock, P.V.E., Stein, N.D. and Stocks, N.G.
1992b, Stochastic resonance for periodically modulated noise intensity.
{\it Phys. Rev. A} {\bf 46}, R1713-1716.

Dykman, M.I., Luchinsky, D.G., Mannella, R., McClintock, P.V.E., Stein, N.D.
and Stocks, N.G. 1993a, Stochastic resonance: linear response and giant
nonlinearity. {\it J. Stat. Phys.} {\bf 70}, 463-478.

Dykman, M.I., Luchinsky, D.G., Mannella, R., McClintock, P.V.E., Stein, N.D.
and Stocks, N.G. 1993b, Nonconventional stochastic resonance.
{\it J. Stat. Phys.} {\bf 70}, 479-499.

Dykman, M.I., Luchinsky, D.G., Mannella, R., McClintock, P.V.E., Stein, N.D.
and Stocks, N.G. 1993c, High frequency stochastic resonance in periodically
driven systems. {\it Sov. Phys. JETP Lett} to be published.

Fauve, S. and Heslot, F. 1983, Stochastic resonance in a bistable system.
{\it Phys. Lett. A} {\bf 97}, 5-7.

Fox, R.F. 1989, Stochastic resonance in a double well. {\it Phys. Rev. A}
{\bf 39}, 4148-4153.

Gammaitoni, L., Marchesoni, F., Menichella-Saetta, E. and Santucci, S.
1989, Stochastic resonance in bistable systems. {\it Phys. Rev. Lett.}
{\bf 62}, 349-352.

Gammaitoni, L., Marchesoni, F., Menichella-Saetta, E. and Santucci, S.
1990, Reply to Comment on stochastic resonance in bistable systems.
{\it Phys. Rev. Lett.} {\bf 65}, 2607.

Gammaitoni, L., Martinelli, M., Pardi, L. and Santucci, S. 1991a, Observation
of stochastic resonance in bistable electron-paramagnetic-resonance systems.
{\it Phys. Rev. Lett.} {\bf 67}, 1799-1802.

Gammaitoni, L., Marchesoni, F., Martinelli, M., Pardi, L. and Santucci, S.
1991b, Phase shifts in bistable EPR systems at stochastic resonance.
{\it Phys. Lett. A} {\bf 158}, 449-452.

Hu Gang, Nicolis, G. and Nicolis, C. 1990, Periodically forced
Fokker-Planck equation and stochastic resonance. {\it Phys. Rev. A}
{\bf 42}, 2030-2041.

Jung, P. and Hanggi, P. 1990, Resonantly driven Brownian motion:
basic concepts and exact results. {\it Phys. Rev. A} {\bf 41}, 2977-2988.

Jung, P. and Hanggi, P. 1991, Amplification of small signals via stochastic
resonance. {\it Phys. Rev. A} {\bf 44}, 8032-8042.

Landau, L.D. and Lifshitz, E.M. 1980 {\it Statistical Physics}, 3rd ed.,
Part 1, Pergamon, New York.

Longtin, A., Bulsara, A. and Moss, F. 1991, Time interval sequences
in bistable systems and the noise-induced transmission of information
by sensory neurons.  {\it Phys. Rev. Lett.} {\bf 67}, 656-659.

McNamara, B., Wiesenfeld, K. and Roy, R. 1988, Observations of stochastic
resonance in a ring laser.  {\it Phys. Rev. Lett.} {\bf 60},
2626-2629.

McNamara, B., and Wiesenfeld, K. 1989, Theory of stochastic resonance.
{\it Phys. Rev. A} {\bf 39}, 4854-4869.

Moss, F. and McClintock, P.V.E., ed. 1989, {\it Noise in Nonlinear Dynamical
Systems}, Cambridge University Press, in 3 vols.

Moss, F., Bulsara A. and Shlesinger, M.F. ed. 1993, Proceedings of the
NATO ARW: Stochastic resonance in Physics and Biology.  {\it J. Stat. Phys.}
{\bf 70}, nos 1/2, special issue.

Nicolis, C. 1982, Stochastic aspects of climatic transitions - response
to a periodic forcing. {\it Tellus} {\bf 34}, 1-9.

Presilla, C., Marchesoni, F. and Gammaitoni, L. 1989. Periodically
time-modulated bistable systems: Nonstationary statistical properties.
{\it Phys. Rev A} {\bf 40}, 2105-2113.

Risken, H. 1989, {\it The Fokker-Planck Equation}, 2nd edn.,
Springer-Verlag, Berlin.

Spano, M.L., Wun-Fogle, M. and Ditto, W.L. 1992,
Experimental observation of stochastic resonance in a magnetoelastic ribbon.
{\it Phys. Rev. A} {\bf 46}, 5253-5256.

Stocks, N.G., Stein, N.D., Soskin, S.M. and McClintock, P.V.E.
1992, Zero-dispersion stochastic resonance. {\it J. Phys. A}
{\bf 25}, L1119-1125.

Stocks, N.G., McClintock, P.V.E. and Soskin, S.M. 1993a,
Observation of zero-dispersion peaks in the fluctuation spectrum
of an underdamped single-well oscillator. {\it Europhys. Lett.} {\bf 21},
395-400.

Stocks, N.G., Stein, N.D. and McClintock, P.V.E. 1993b, Stochastic
resonance in monostable systems, {\it J. Phys A} {\bf 26}, L385-390.

Thompson, J.M.T. and Stewart H.B. 1986, {\it Nonlinear Dynamics and
Chaos}, Wiley, New York.

Wiesenfeld, K. 1993, Signals from noise: stochastic resonance pays off,
{\it Physics World} {\bf 6}, February pp 23-24.

\newpage
\centerline {\bf Figure Captions}

\begin{itemize}
\item[1.] Signal/noise ratio $\tilde R$ for the electronic
model of the double-well oscillator (6), as a function of scaled noise
intensity $T/\Delta U$.  Direct measurements of $\tilde R$ (squares)
are compared with results calculated from measured spectra $Q^{(0)} (\omega)$
(crosses) using the fluctuation dissipation relations (5).  (After Dykman et
al, 1990a).

\item[2.] The phase shift $-\phi$ (degrees) between the periodic force and
the response measured (data points) for an electronic model of the overdamped
double-well system (7) with $\Omega$ = 0.1 and $A$ = 0.04 (circles)
and $A$ = 0.2 (squares).  The full curve represents the LRT theory
based on (5); the dashed curve represents the prediction of earlier
2-state theories.  Inset: the normalised signal/noise ratio as a
function of noise intensity $D$, showing that the LRT theory works well near
the minimum. (After Dykman et al, 1990a).

\item[3.] The signal/noise ratio $R$ of the response of the system (9) to a
weak trial force at frequency $\Omega$, as a function of noise intensity
$\alpha$, in experiment and LRT theory: at the trial frequency $\Omega$
(circle data and associated curve); and at the mirror-reflected frequency
$\vert 2\omega_F - \Omega \vert$ (squares).  For noise intensities near and
beyond the maxima in $R (\alpha)$, the asymptotic theory is only qualitative
and so the curves are shown dotted. After Dykman et al, 1990c).

\item[4.] (a) The squared stochastic amplification factor $S^2$ measured
for the electronic model of the system (10) with $\vert B \vert$ = 2,
$A$ = 0.02, (circle data) compared with the LRT theory (curve) based
on the fluctuation dissipation relations (5).  (After Stocks et al (1993b)
where the square data points and associated curve, for $B$ = 0, are also
discussed.)
\end{itemize}
\end{document}